\documentclass[referee,a4paper,12pt,traditabstract]{jswsc} 

\usepackage{graphicx}
\usepackage{txfonts}
\usepackage{subfigure}
\usepackage{epstopdf}
\usepackage[displaymath,mathlines]{lineno}
\usepackage[authoryear,round]{natbib}
\usepackage[backref]{hyperref}
\usepackage{url}

\usepackage{amsmath}
\usepackage{siunitx}
\hypersetup{colorlinks=true,citecolor=cyan,urlcolor=cyan,linkcolor=blue}

\begin{document}


   \title{Performance of radio-based detection to operational monitoring M5+ class solar flares}

   
   \titlerunning{Radio-based solar flare detection}

   \authorrunning{van Ravenswaaij et al.}

   \author{J. van Ravenswaaij
          \inst{1}
          \and
          I. Wilms\inst{1}
          \and
          I. Ricardo\inst{1}
          \and 
          M. Brchnelova\inst{2}\fnmsep\thanks{Corresponding author}
          }

   \institute{Department of Quantitative Economics, Maastricht University, \\ 
              \email{\href{mailto:i.wilms@maastrichtuniversity.nl}{i.wilms@maastrichtuniversity.nl}}
         \and
             Faculty of Military Sciences, Netherlands Defence Academy, Den Helder \\
             \email{\href{mailto:m.brchnelova@mindef.nl}{m.brchnelova@mindef.nl}}
             }


  \abstract
   {Early detection of major solar flares is critical for defense operations due to their potential to disturb radar and radio systems. Typically, soft X-ray flux is used to monitor and classify solar flares, but since this flux has to be measured in space, it means that its availability itself is dependent on space weather conditions. For this reason, in this paper, we investigated the feasibility of using ground radio observations to monitor major (M5+ class) solar flares. We made use of datasets from the GOES-16 satellite and the Radio Solar Telescope Network in the time range between March 2023 and March 2025. An elastic net regularized logistic regression model was trained on this data, optimized through a grid search and with incorporated class weighting for class imbalance.  It was found that especially higher frequencies (8800 MHz) had a reasonable ability in monitoring and  predicting major flares (
   {precision and recall for flare events are 53\% and 65\%, respectively--  implying that roughly one third of flares were not detected --}with signals appearing, on average, {3 to 4} 
   minutes before the M5 threshold is exceeded). {Radio} measurements at super high frequencies can thus serve as an alternative method to monitor major solar flaring activity.}


   \maketitle



\section{Introduction}
\label{sec:int}
Strong solar flares can affect radio and radar systems in two distinct ways. Firstly, communication via the high-frequency (HF) radio channel can be disturbed if the electromagnetic radiation from a flare sufficiently enhances the D-layer ionization in the ionosphere, causing excessive absorption and anomalous refraction of HF signals and preventing these from reaching the higher layers normally used for propagation. Secondly, the very-high frequency (VHF), ultra-high frequency (UHF) and super-high frequency (SHF) communication may be disturbed by interference of strong solar radio bursts, SRBs, associated with solar flares (as well as other space weather phenomena), which may increase the noise background to the point of a complete blackout. For this reason, monitoring solar flares-- the focus of this paper --is important for many users of radio and radar systems. 

Currently, solar flares are monitored and classified according to the measured soft X-ray flux from the Sun, for example by the GOES X-ray sensor (XRS) \citep{Woods2024}. Since the X-ray radiation is absorbed by the atmosphere, this measurement cannot be conducted on the ground. This, however, means that solar flare monitoring and classification is dependent on space infrastructure, which itself is vulnerable and can become unavailable during major space weather events. For this reason, it is essential to start developing also other, more robust methods of (near)real-time monitoring and classification of solar flares. Since major solar flares are also often observable in the radio spectrum as (predominantly type III) solar radio bursts, SRBs, (see papers going all the way back to \citealp{Loughhead1957}) investigating the detailed association between space-based X-ray data and ground-based solar radio measurements is a logical first step towards this objective.

{Another reason for our increased interest in solar flare signatures in radio is that these could be exploited operationally with systems such as the Dutch DISTURB (Disturbance-detection by Intelligent Solar radio Telescope of (Un)perturbed Radiofrequency Bands) \citep{DISTURB}. Once operational, this system could issue automatic warnings about solar radio bursts to the relevant VHF (very-high frequency), UHF (ultra-high frequency) and SHF (super-high frequency) radio and radar end-users. However, if a statistically significant connection is found between certain radio frequencies and the occurrence of M5+ class solar flares (as measured in X-ray), the system could also be used to generate warnings for long-distance HF (high-frequency) radio users.}

During a solar flare, different processes produce a variety of emissions at various energies. Soft X-ray flux (typically up to ten keV), such as what is measured by GOES, is typically predominantly associated with thermal radiation and thus can be used to interpret plasma heating during the flaring process (see, for instance, \citealp{Garcia1998} or \citealp{Mithun2022} for a more recent study). Hard X-ray flux (tens of keV), in contrast, is associated with bremsstrahlung by high-energy electrons. \citet{Isola2007}, however, found a strong correlation between the soft and hard X-ray flux despite their different origins. 

The same high-energy electron population that causes bremsstrahlung and thus hard X-ray is thought to cause gyrosynchrotron emission, producing microwave radiation at the same time (at centimeter to decimeter wavelengths, thus roughly 3 to 30 GHz {though even frequencies higher than that have been observed \citep{Wu2024, Xu2025,Trottet1986,White2011})}, which also results in the fact that these two emissions have similar profiles \citep{Dennis1988}. Thus, it is reasonable to expect that an increased soft X-ray flux from a solar flare would be, to some degree, also correlated with an increased microwave emission. This is, for example, shown in the work of \citet{Keitarou2023}, where the microwave peaks and soft X-ray peaks are compared with each other. At lower energies in the radio spectrum (decimeter to meter wavelengths, thus roughly 0.3 to 3 GHz), radio emission likely does not originate from single electrons, but from plasma waves \citep{Melrose1986,Li2014,Ratcliffe2014}. Still, despite the different processes causing these emissions, there seems to be a correlation between the two \citep{Shamsutdinova2024}. For more discussion about the different forms of emissions and the corresponding processes, we refer the reader to the reviews of \citet{Bastian1998} and \citet{Benz2017}. It thus becomes of interest to investigate to what extent these different radio signatures can predict some aspects of evolution of the soft X-ray flux that we use currently for classification of solar flares. 

Some of the early investigations into the relationship between solar X-ray emission and accompanying radio signatures were those of \citet{Kane1981}. Through comparing radio signals with hard X-ray (10–100 keV), they found that fewer than 20\% of all X-ray flares showed detectable radio signatures. However, when considering only the stronger solar flares, such as the ones relevant for our study, the association rate rose sharply to 70–80\%. A similar figure, 83\%, was also found by \citet{Benz2005} when comparing soft X-rays ($<$10 keV) and radio measurements for flares of classes C5+, indicating that the strength is indeed an important factor and that for strong flares, radio signatures can be expected. 

The work of \citet{Giersch2017}, who investigated solar radio signatures from the period of 34 years, indicates that they are more likely to occur at lower frequencies (below 1 GHz). This is generally expected, since lower-energy processes are generally more prevalent. However, this analysis was conducted based on all the measured radio signatures, without connecting these to the specific classes of space weather events, meaning that many of these could have been connected to other space weather phenomena such as coronal mass ejections. A thorough investigation of radio and hard X-ray emissions of weaker events, mostly B and C class flares, was also conducted by \citet{Reid2017}. Strength of the event and the measured X-ray channel are, however, important aspects from an operational standpoint and as shown in the previous paragraph, the statistics across the various classes of flares and X-ray channels may differ. 

For this reason, in this paper, we focus only on the statistics of the stronger solar flares of M5+ to determine whether their radio signatures could be a good enough predictor of their occurrence. Specifically, 
we use logistic regression to estimate  the likelihood of an X-ray solar flare occurring based on  data from several radio channels.  Intuitively, we are
 interested in knowing how likely it is, if we observe an increase in intensity in a specific radio channel, that this corresponds to an M5+ class solar flare. 
Since radio signatures occur at a wide variety of radio channels, we are also interested in determining which radio channel is the most suitable to be monitored for this purpose. To this end, we use an elastic net regularized logistic regression model that selects the most relevant radio channels for detecting an M5+ class flare. Since this study is focused on operations, we further make no distinction between the types of radio bursts or processes causing signatures at higher versus lower radio wavelengths. In practice, what we observe real-time is an increase in intensity in a particular radio channel and that is in principle all the information that we have at the time to predict whether an M5+ class flare is currently occurring. 
{Furthermore, since the impact of the flaring event on long-distance radio is highly dependent on the X-ray intensity and since solar flares are currently operationally detected and classified according to their X-ray intensity, in this work, we do not take into account flares that do not show strong (M5+) X-ray signatures, such as weaker flares and microflares.}
Section \ref{sec:met} details the preparation of {radio} and X-ray datasets and further describes the logistic regression model using elastic net regularization, hyperparameter tuning, and correction for unbalancedness. Section \ref{sec:res} presents empirical results, highlighting key predictive channels and providing individual frequency analysis for the eight channels. Section \ref{sec:dis} discusses the physical interpretation, limitations and recommendations, followed by conclusions in Section \ref{sec:con}.

\section{Methodology}
\label{sec:met}

\subsection{Data sources}

The purpose of this work is to investigate the connection between signatures observed in radio observations and strong solar flares, here defined as class M5+ as measured in soft X-ray. Thus we require two types of time-series: radio in different channels and soft X-ray. The events investigated here are from the period around the current solar cycle (SC25) maximum, from 1 March 2023 to 22 March 2025, as this cycle has been more active than the previous one.

The solar flare classification is currently carried out by the long (1–0.8 nm) channel of the X-Ray Sensor (XRS) \citep{Woods2024} of GOES (Geostationary Operational Environmental Satellites). As we are interested in solar flares classified according to this instrument, we used GOES-16 soft X-ray data in our study. In the given period, GOES-16 had an almost complete coverage. 

For radio analysis, we used data of the Radio Solar Telescope Network (RSTN) established by the United States Air Force Research Laboratory (for technical details, see e.g. \citet{Giersch2022}). The reason for this was the fact that it has stations all over the world and thus almost a complete coverage of the Sun, with instruments that have the same radio channels and similar specifications, making data fusion possible.  These stations are in San Vito (45$^\circ$ 55' N / 12$^\circ$  52' E), Sagamore Hill (40$^\circ$  53' N / 73$^\circ$  30' W ) and Pālehua (22$^\circ$  03' N / 159$^\circ$  31' W). Each station records radio flux at 8 different channels, see Table \ref{tab:channels}. A two-year long data set spanning March 2023 to March 2025 was selected to ensure a sufficient number of solar flares in this period. 

\begin{table}[t]
\centering
\caption{Radio frequencies observed by the RSTN in San Vito, Sagamore Hill and Pālehua.}
\label{tab:channels}
\begin{tabular}{cc}
\hline
\textbf{Channel} & \textbf{Frequency (MHz)} \\
\hline
1 & 245     \\
2 & 410     \\
3 & 610     \\
4 & 1415    \\
5 & 2695    \\
6 & 4995    \\
7 & 8800    \\
8 & 15400   \\
\hline 
\end{tabular}
\end{table}

\subsection{Data} \label{subsec:data}

To select appropriate statistical methodology for this work, 
we first explore the nature of the selected data. The classification of all the detected flares and the selection of M5 and above are given in Appendix in Figure \ref{fig:flares_all_Xray}. 
Over the course of these two years, 178 flares out of the total of 924 flares exceeded the M5 threshold ($5\cdot10^{-5}$ W/m$^2$).

Figure \ref{fig:flares_per_day} in the Appendix shows the number of recorded M5+ events per day, with an ``event" being  the occurrence of a flare exceeding the M5 threshold. On days with at least one M5+ flare (active days), the mean count is 1.51 flares, so most active days see either one or two events, with more than two being rare. The largest solar flare count is six solar flares in a single day, which occurred in late May 2024.

Since our work is application-focused, we also measure the duration of a flare in a manner different than what is used by NOAA (The National Oceanic and Atmospheric Administration). As we are interested in X-ray flux higher than the M5 threshold (as this creates, for instance, a specific measurable level of absorption loss in an HF communication link), we further define the duration of a flare in this study as the duration of time that its X-ray flux stayed above this level. 
Figure \ref{fig:flare_duration} in the Appendix displays the duration of the flares. The mean of 21.21 minutes is sensitive to the few extraordinarily long events. The median is 10 minutes and the mode 1 minute, showing that most of the flares just touched the M5 threshold before dimming under it, as would be also expected.

\begin{figure}
    \centering
    \includegraphics[width=1\linewidth]{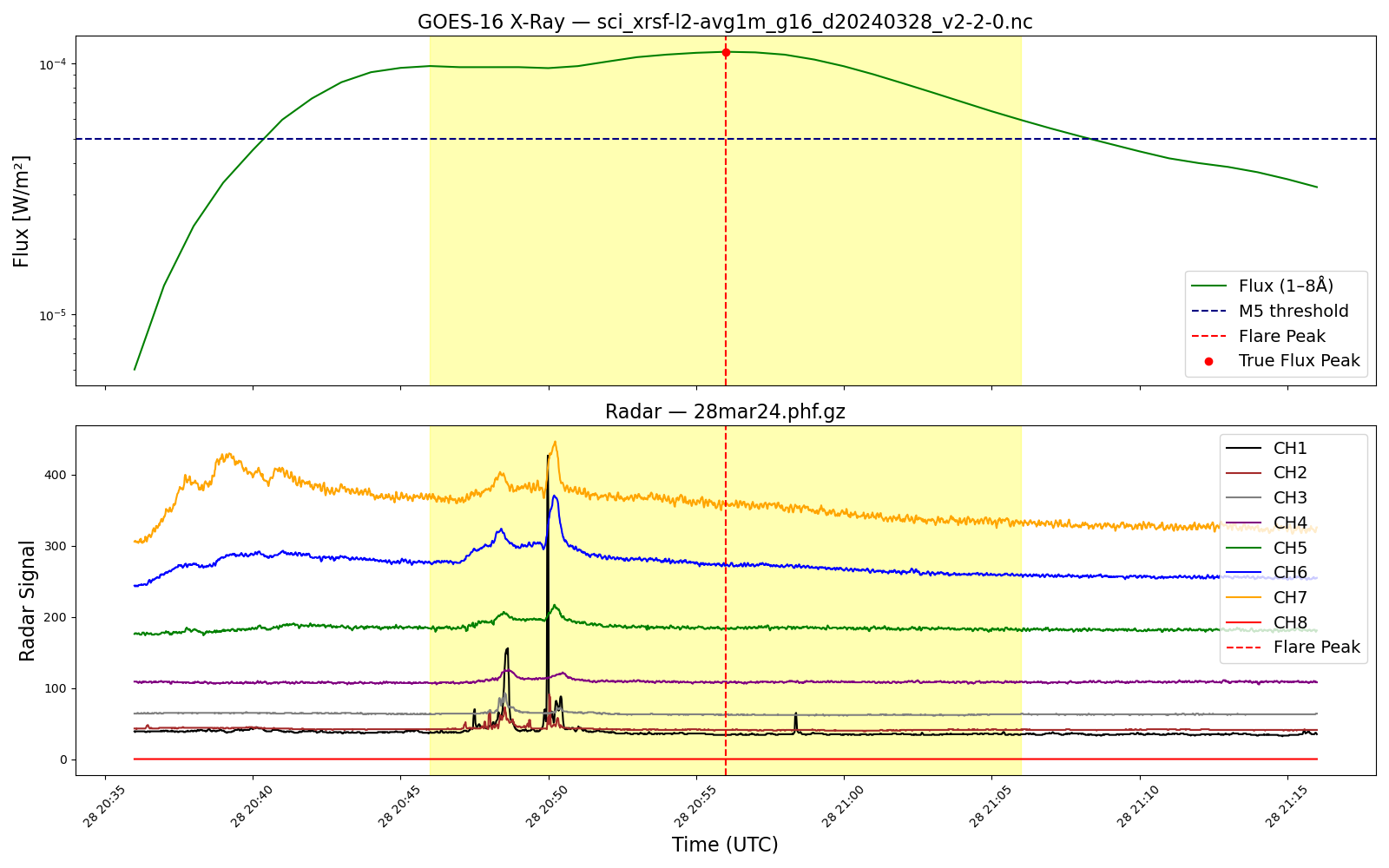}
    \caption{X-ray and corresponding {radio} data during a solar flare at  28 March 2024}
    \label{fig:flare_XrayRadioexmaple}
\end{figure}

An example comparison between a X-ray and radio time series during a flare is shown in Figure \ref{fig:flare_XrayRadioexmaple}. On the top, the soft X-ray data, measured every minute, is shown and on the bottom the different radio channels as measured every second by RSTN. 
In the remainder, to address the temporal resolution mismatch of the RSTN data (seconds) and the X-ray data (minutes), we transformed the radio data to one-minute interval time series by taking the maximum value within each minute to preserve significant signal amplifications.
The red line in Figure \ref{fig:flare_XrayRadioexmaple} indicates the peak of the X-ray flux. One of the 
patterns observed during this work was that the radio signals would almost consistently appear well before X-ray flux exceeded the given threshold. For that reason, in this study, we did not only include  
the  original one-minute radio time series, but also ``lagged" radio time series as predictors in the model. These lagged time series effectively represent earlier values of the signal. For example, a lag of 1 corresponds to the radio observation 1 minute before the current point and so on. By including up to ten lags (equivalent to 10 minutes), the model can assess how earlier patterns in the radio time-series influence its predictive relationship with later X-ray observations.

Starting with the 178 M5+ flares, flares  for which radio or X-ray data were not available, were too short or seemed to have been corrupted were removed, resulting in 116 solar 
flares up for analysis, see Figure \ref{fig:figure data selectio} for an overview of the data preparation steps. For every flare, we retrieved 
an X-ray and radio time series window whose length was determined by the start and end point of the station providing the radio data during the full duration of the solar flare. If multiple M5+ flares occurred during the same window, we kept all solar flares but removed duplicate one-minute observations. This resulted in 
129  windows with M5+ solar flare activity, meaning that at least one M5+ flare occurred during the recorded window, and around 458  one-minute observations (roughly eight hours) per window.\footnote{Note that the number of windows (129) is larger than the number of solar flares (117) since in some cases, the solar flare is observed by two stations. In such cases, we constructed the time windows for both stations.} 
To jointly analyze the predictive power of radio signature data for M5+ solar flare occurrence, we stacked all time windows with M5+ solar flare activity, thereby preserving the chronological order, which resulted in a final single dataset containing
59,162  one-minute observations {of which 3,438 correspond to solar
flares (5.81\%)}. 


\subsection{Statistical methodology}
To estimate the likelihood of an M5+ solar flare occurring, based on radio signature data, we use a logistic regression model. 
Since the occurrence of a flare is a binary outcome, a logistic regression model is appropriate for estimating the probability of its occurrence from the radio predictor variables.
Furthermore, we aim to identify the radio channels that contain most predictive power for the occurrence of an M5+ solar flare. 
To identify the most relevant radio channels, we use a logistic regression model with elastic net regularization. 
For details on the statistical methodology, we refer the interested readers to \cite{hastie2009elements}, Chapter 4.

\subsubsection{Logistic regression}
Let $y_t$ denote the response variable that takes on the value one if an M5+ solar flare occurs (i.e.\ M5 threshold exceeded) during minute $t$, and zero otherwise.
Let $x_t=(x_{1,t}, \ldots, x_{p,t})^\top$ denote the $p$-dimensional  vector of radio signature variables at time $t$. In our study, we use eight different radio channels, hence $p=8$.  
The logistic regression model is given by
\begin{equation}
\text{logit}(\pi_t) = \text{ln}\left( \frac{\pi_t}{1- \pi_t}\right) = \mu_0  + \sum_{i=1}^p\sum_{j=0}^\ell \beta_{i,j} x_{i,t-j}, \label{eq:LRM} 
\end{equation}
where ln($\cdot$) denotes the natural logarithm, 
$\pi_t$ the probability of an M5+ solar flare occurring at time $t$ given the radio signature predictors,
and logit (aka log-odds) the odds of a flare occurring in natural logarithm form. Furthermore,
$x_{i,t-j}$ is the $j$th lag of the $i$th radio channel, where the original radio channel time series $x_{i,t}$ is obtained for lag $j=0$, and we include up to $\ell=10$ lags in the logistic regression model, see Section \ref{subsec:data}.  The parameters that need to be estimated are  $\mu_0$, denoting the intercept, and the $\beta_{i,j}$s representing the coefficients of radio channel $i$ at lag $j$.

Unlike the standard regression model, the logistic regression model in \eqref{eq:LRM} maps any input to $\pi_t = 1/(1 + e^{-z_t})$, with $z_t = \text{logit}(\pi_t)$
where $\pi_t$ thus takes on any value between 0 and 1 through the usage of the logistic function, making it appropriate for modeling the probability of an M5+ solar flare occurring. 
In particular, one typically uses the threshold 0.5 to predict a  flare occurring at minute $t$ if  $\hat{\pi}_t \geq 0.5$, and no solar flare occurring otherwise.

To interpret the parameters in the logistic regression model, note that 
$\mu_0$ gives the ``baseline" log-odds
of a flare when all predictors are zero.
Furthermore, $\beta_{i,j}$ gives the change in the log-odds of the response for a one-unit increase in the corresponding predictor $x_{i,j}$ (ceteris paribus, which we suppress in the remainder for compactness).
A more intuitive interpretation is obtained by exponentiation of the coefficient, $e^{\beta_{i,j}}$, resulting in the odds ratio (OR) that represents the multiplicative change in the odds of a flare occurring for a one-unit increase in the predictor. 
An OR larger (smaller) than one indicates that the occurrence of a flare becomes more (less) likely as $x_{i,t-j}$ increases; there is no effect if the OR equals one.

\subsubsection{Elastic net regularized logistic regression}
The logistic regression model in equation \eqref{eq:LRM} contains a large number of parameters that need to be estimated namely: 1 ({$\mu_0$}) + $8\times 11$ ($\beta_{ij}$s) = 89 parameters.
Furthermore, jointly with parameter estimation, we aim to identify/select the most relevant radio signature variables for modeling the likelihood of a solar flare occurring. 
To ensure accurate estimation of the many parameters and selection of the most suitable channels, we resort to elastic net regularization  \citep{ZouHastie2005} of the logistic regression model.

The elastic net estimates are obtained by maximizing the  penalized likelihood given by
\begin{equation}
 \sum_{t=1}^{n} \left[ y_t \log(\pi_t) + (1 - y_t) \log(1 - \pi_t) \right] - \lambda \left[ \alpha \|\beta\|_1 + (1 - \alpha) \|\beta\|_2^2 \right], \label{eq:elasnet}
\end{equation}
where $n$ denotes the sample size, $\beta$ the $p\ell$-dimensional parameter vector collecting all coefficients corresponding to the radio signature predictors (i.e.\ all $\beta_{ij}$s). To encourage selection of the most suitable radio channels, a penalty term is subtracted from the log-likelihood in equation \eqref{eq:elasnet}. The elastic net penalty consists of a convex combination of an $\ell_1$-penalty ($\|\beta\|_1$) and an $\ell_2$-penalty ($\|\beta\|_2$) with weight $\alpha \in [0,1]$. 
One could opt for mere lasso penalization ($\alpha=1$), but we allow for
$0 \leq \alpha \leq 1$, 
since the addition of the $\ell_2$-penalty helps to address the collinearity among the lagged radio signature variables.

The tuning parameter $\lambda>0$ regulates the overall strength of the penalization with $\lambda=0$ no penalization and more penalization when $\lambda$ increases. The larger $\lambda$, the sparser the estimates, meaning that more of the $\beta_{ij}$s will be estimated at zero. As commonly done in the regularization literature, we first standardize all radio signature variables (to have mean zero and standard deviation one) such that all variables are equally penalized regardless of their original units. 

In the following, we fix  $\alpha=0.5$ (to reduce computational burden), which corresponds to an equal balance between the $\ell_1$ and $\ell_2$ penalties, and tune the parameter $\lambda$ using a cross-validation procedure that accounts both for the time series nature of the data and the class imbalance of the solar flare (in)activity. The procedure is detailed in Appendix \ref{app:lambda}. 
As cross-validation score, we use the $F_1$ score, a popular evaluation metric for logistic regression because it balances precision and recall. This makes it especially useful when classes are imbalanced, as is the case in our study. The score is given by
\begin{equation}
F_{1} = 2 \times \frac{\text{Precision} \times \text{Recall}}{\text{Precision} + \text{Recall}}, \quad
\text{Precision} = \frac{\text{TP}}{\text{TP} + \text{FP}}, \quad
\text{Recall} = \frac{\text{TP}}{\text{TP} + \text{FN}}, \nonumber
\end{equation}
{where 
    TP denotes the True Positives, that is, the number of actual solar flare ``events" correctly predicted by the model,
    TN denotes the True Negatives, that is, the number of non-flare events correctly predicted by the model,
    FP denotes the False Positives, that is, the number of non-flare events that the model incorrectly classified as flares, and 
    FN denotes the False Negatives, that is, the  number of actual solar flare events that the model failed to detect.
Precision measures the proportion of True Positive predictions amongst all positive predictions while Recall (also called ``Sensitivity") is the proportion of True Positives amongst all actual positives.
}
The higher the $F_1$ score, the better since it indicates both high precision and recall.

\section{Results}
\label{sec:res}
We present the results of the elastic net regularized logistic regression model with tuned hyperparameter $\lambda = 2\,836$, as obtained based on the time-series cross-validation procedure, see Appendix \ref{app:lambda:results} for details.

\subsection{The estimated logistic regression model}

The estimated model is given by
\begin{equation}
\widehat{\text{logit}}(\pi_t)= -0.2794  + 0.0101 \cdot \text{ch8}_t + \sum_{j=0}^{10} \hat{\beta}_{7,j} \cdot \text{ch7}_{t - j},
\end{equation}
where $\text{ch7}_t$,  $\text{ch8}_t$ denote the time series for channels 7 and 8 respectively and the instantaneous and lagged coefficient values for channel 7 are given by
\begin{multline*}
{\hat{\beta}_{7}}= [\ 0.0532,\ 0.1217,\ 0.1387,\ 0.1479,\ 0.1550,\ 0.1396, \\
0.1194,\ 0.1069,\ 0.0809,\ 0.0661,\ 0.1096\ ].
\end{multline*}
The elastic net regularized regression model, at the optimally tuned $\lambda$, thus reduced the model complexity by shrinking 76 out of 88 radio signature predictors to zero; as can also be seen from the red line in the bottom panel of Figure \ref{fig:hyperparametertuning} in Appendix \ref{app:lambda:results}. 
Only 12 radio predictors are retained. In particular, all channel 7 predictors (the instantaneous one and the lags) are retained in the model along with the instantaneous value of channel 8, indicating that channel 7 is the best predictor for M5+ solar flare occurrence.
{To provide reliable inference in the presence of variable selection, $p$-values are computed and reported in Table \ref{tab:logistic-impact} (column ``$p$-value") using the post-double selection procedure \citep{belloni2017program}; see \cite{belloni2016post} for further details on post-double-selection for generalised linear models including the logistic regression model. 
The
contemporaneous values of channel 8 and channel 7 (i.e., $ch8_t$ and $ch7_t$) are found to be statistically insignificant (at the 5\% significance level), all other variables are statistically significant, indicating their strong
predictive relevance.
}

Table \ref{tab:logistic-impact} further presents the impact of the coefficients on the odds and the probability of a solar flare occurring. 
In particular, the column ``Channel" lists the channels with non-zero estimated coefficients  
$\hat{\beta}_{i,j}$s in the column ``Coefficient", and the  corresponding odds ratio $e^{\hat{\beta}_{i,j}}$ is displayed in the column ``Odds Ratio". All ORs are above one indicating that the occurrence of a solar flare becomes more likely as the intensity in the channels increases. 
For channel 8, for instance, a one-unit increase in its intensity, increases the log-odds by 0.0101 and hence the odds of a flare by 1\%.
By accounting for the baseline odds ($e^{\hat{\mu}_0} = e^{-0.2794} = 0.7562$), one can then compute the increase in predicted probability (column ``Change in Predicted Probability") by 
\begin{equation}
\frac{e^{\hat{\mu}_0}e^{\hat{\beta}_{i,j}}}{1+ e^{\hat{\mu}_0}e^{\hat{\beta}_{i,j}}} - \frac{e^{\hat{\mu}_0}}{(1+ e^{\hat{\mu}_0})},  \nonumber
\end{equation}
which amounts to
\[
\frac{e^{-0.2794}\,e^{0.0101}}{1 + e^{-0.2794}\,e^{0.0101}}
\;-\;\frac{e^{-0.2794}}{1 + e^{-0.2794}}
\approx 0.002450 = 0.25\%
\]
for channel 8. 
Thus, to reach the 50\% decision threshold (\(\hat{\pi}_{t}\ge0.5\)) that an M5+ solar flare is taking place solely by variations in a particular predictor, the signal at this channel must increase by $-\hat{\mu}_0/\hat{\beta}_{i,j}$ which amounts to $0.2794/0.0101 \approx 28$ units for mere channel 8 variations, see column ``Required Unit Change".

In contrast, the four-minutes lagged channel 7 ($ch7_{t-4}$) produces a much larger change per unit. In particular, a one-unit increase in channel 7 increases, four minutes later, the log-odds by 0.1550, and hence the odds of a flare by 17\%. A one unit-increase in intensity at channel 7, results, four minutes later, in an increased predicted probability of 3.83\% and only 1.80 units are required to lift the probability of a solar flare occurring four minutes later above 50\%.

\begin{table}[t]
\color{black}
  \centering
  \caption{Logistic regression coefficient impact analysis.}
  \label{tab:logistic-impact}
  \begin{tabular}{l S S S S S}
    \hline
    Channel     & {Coefficient} & {$p$-value}  & {Odds Ratio} & {Change in Predicted} & {Required }\\
         & {} & {}  & {} & { Probability} & {Unit Change}\\
    \hline
    $ch8_t$& 0.0101  & 0.3289  & 1.0102	& 0.0025	& 27.6634           \\
    $ch7_t$& 0.0532   &   0.1328    & 1.0546 &	0.0131 & 
	5.2519            \\
    $ch7_{t-1}$& 0.1217  &   0.0337      & 1.1294 &
	0.0301 & 2.2958 \\
    $ch7_{t-2}$& 0.1387  &  0.0303       & 1.1488 & 0.0343 & 2.0144 \\
    $ch7_{t-3}$& 0.1479  &   0.0117      & 1.1594 & 0.0366 & 1.8891 \\
    $ch7_{t-4}$& 0.1550  &    0.0024     & 1.1677 & 0.0383 & 1.8025
           \\
    $ch7_{t-5}$& 0.1396  &  0.0014       & 1.1498 & 0.0345 & 2.0014
           \\
    $ch7_{t-6}$& 0.1194  &   0.0031      & 1.1268 & 0.0295 & 2.3400
           \\
   $ch7_{t-7}$& 0.1069  &   0.0042     & 1.1128 & 0.0264 & 2.6137
            \\
    $ch7_{t-8}$& 0.0809  &  0.0114      & 1.0843 & 0.0199 & 3.4536
           \\
    $ch7_{t-9}$& 0.0661  &  0.0111       & 1.0683 & 0.016 & 4.2269
           \\
    $ch7_{t-10}$& 0.1096  &   0.0047      & 1.1158 & 0.0271 & 2.5493
            \\
    \hline
  \end{tabular}
\end{table}

\subsection{Confusion matrix and classification report}
The performance of the model is, of course, not 100\% accurate. The confusion matrix, showing the model's performance 
by comparing predicted classifications against actual outcomes, can be found in Table \ref{tab:confusion-matrix}. 
It presents the performance of the elastic net regularized logistic regression model on the test data, namely the last approximately 30\% of the time points.
{Note that the entire dataset contains 59,162 one-minute observations, of which 3,438 correspond to solar flares (5.81\%). 
The first 41,252 observations (training and validation set) contain 2,092 flare events (5.07\%), while the remaining 17,910 observations (test set) contain 1,346 flare events (7.52\%), see Appendix \ref{app:lambda} for further details.
At the flare level, the training and validation set contain 85 out of the 115 flares, the test set contains 30.
}

Out of 16\,564 one-minute time points without flares, 15\,789 were correctly classified (True Negatives, TN), and 775 were incorrectly flagged as flares (False Positives, FP). Of the 1\,346 one-minute time points with actual flares (at the one-minute level), 
877 were correctly recovered (True Positives, TP) and 469 were missed (False Negatives, FN). 

\begin{table}[t]
  \centering
  \caption{Confusion matrix of the elastic net regularized logistic regression model.}
  \label{tab:confusion-matrix}
  \begin{tabular}{lcc}
    \hline
      & \textbf{No Predicted Flare}& \textbf{Predicted Flare}\\
    \hline
    \textbf{Actual: No Flare}& 15\,789 (TN)& 775 (FP)\\
    \textbf{Actual: Flare}& 469 (FN)& 877 (TP)\\
    \hline
  \end{tabular}
\end{table}

The corresponding classification report is given in Table \ref{tab:classification-report}, split according to the two classes where class 0 contains the  16\, 564 one-minute time points without a solar flare and class 1 the 1\,346 one-minute time points with a solar flare.
{The class of interest is the ``True" flare class as our primary interest lies in correctly identifying flare events rather than true negatives. We therefore focus on the precision and recall of the True (flare) class.
The precision for the ``Flare" class of 0.5309 and its recall of 0.6516 indicate that the model  detects roughly two-thirds of actual flares (at the one-minute level) . The combined $F_1$-score of 0.5851 reflects this imbalance. 
For observations conducted under such conditions (e.g., ground-based radio observations), the precision and recall values are reasonable.}

\textcolor{black}{
Furthermore, while the confusion matrix and classification performance are reported at the one-minute level, it is also informative to evaluate performance at the flare level itself, specifically in terms of how many flares were forecast and with what lead time. We define a flare as successfully forecast  if at least one minute within its occurrence window is predicted as positive. Conversely, a flare is missed if no positive minute-level  prediction occurs within this detection window. In addition, we track ``near misses,” defined as cases in which the model predicts an M5+ flare over consecutive minutes within at most 30 minutes before the onset or after the end of an actual M5+ flare, as well as ``false alarms,” defined as cases in which the model predicts an M5+ flare over consecutive minutes at least 30 minutes before or after the start of a flare.}

\textcolor{black}{Of the 30 flares in the test set, 17 were correctly forecast, while 13 were missed and 4 false alarms were recorded. 
These results correspond to a precision of 0.8095,  a recall of 0.5667 and an $F_1$-score of 0.6667. 
Among the 13 missed flares, 2 were classified as near misses, indicating potential for future research to improve the temporal accuracy of flare detection.
Amongst the 17 correctly detected flares,  7 were identified prior to the actual onset, with a median lead time of 4 minutes (meaning that the first one-minute level prediction occurred 4 minutes before the flare onset). A further 5 were predicted to start exactly at the observed onset. The remaining 5 flares were detected with a delay, with the predicted start times occurring a median of 2 minutes after the true onset across these events. Importantly, even in cases where the predicted onset coincides with or follows the true onset, the most pronounced radio signal typically occurs about 4 minutes earlier. In practical operational settings, this still provides a (small) window for automated early warning.
}


\begin{table}[t]
  \centering
  \caption{Classification report of the elastic net regularized logistic regression model.}
  \label{tab:classification-report}
  \begin{tabular}{lrrrr}
    \hline
    \textbf{Class} & \textbf{Precision} & \textbf{Recall} & \textbf{F1-score} & \textbf{Support} \\
    \hline
    0 (No flare) & 0.9712 & 0.9532 & 0.9621 & 16\,564 \\
    1 (Flare) & 0.5309 & 0.6516 & 0.5851 & 1\,346 \\
    \hline
    \textbf{Accuracy} & & & \textbf{0.9305}& \textbf{17\,910} \\
    \hline
  \end{tabular}
\end{table}

\subsection{False negatives and false positives}

A total of 468 observations were flagged as false negatives, meaning that an M5+ class solar flare did occur but the model incorrectly labeled the observation as insufficiently likely for a solar flare to occur (i.e.\ $\hat{\pi}_t < 0.5)$. These 468 one-minute observations correspond to 30 unique solar flares. In three of these, the solar flare was correctly observed by another {radio} station at the same time, meaning that a better data fusion technique and having more stations available could at least partly help to improve detection accuracy and overcome the limitations of the separate stations. In another 14 cases, it was the timing of the prediction that failed. For instance, the X-ray flare appeared too early (e.g., before the radio) or too late. This indicates that the model could possibly be further improved by adjusting the lag structure in the model.

In the rest of the cases, there was no visible signature in the radio channels. {It should be noted that all the events where radio data were not available were already filtered out from the analysis at the very beginning, hence the incomplete radio coverage should not have any impact on this statistics}.  It is known that sometimes even strong flares do not produce observable radio signatures, e.g.\ when these are located at the limb and thus difficult to observe radially \citep{Benz2007}. Equally as likely is the scenario that the signature could have occurred at frequencies not investigated in this study (e.g., between the investigated frequency channels). The latter could be improved by monitoring more radio channels.

The model also produced a total of 775 false positives (a solar flare prediction when none occurred). These predictions were concentrated around 24 distinct solar flares. Ten of these were caused by {radio} signatures that occurred well after the end of the flare in X-ray. For one of these, for instance, the flare-like {radio} activity persisted for over 3 hours and 41 minutes after the X-ray signal dipped below the threshold. These were especially associated with stronger flares. In the case of another 11 flares, the radio signal arrived too early.

For both cases, increasing the model window could improve the accuracy, though it is difficult to tell by just how much this window would have to be increased to sufficiently increase the accuracy without making it too complex and prone to noise. This, however, also shows that in most cases, a strong solar flare was indeed present around the time the prediction was made, which is still valuable from an operational perspective.

\subsection{Key predictors}
Among the retained variables from the elastic net regularized logistic regression model, channel 7 stands out as the dominant one, with the most influential coefficients being $ch7_{t-4}$ (0.1550) and 
{$ch7_{t-3}$ (0.1479)}
This means that channel 7 was particularly successful at predicting solar flares {exploiting radio signals appearing 3 to 4} minutes before the X-ray signal exceeded the M5 threshold, which also implies that channel 7 may contain predictive power to provide a warning a few minutes before the M5 threshold is reached. 

The fact that only channels 8 and 7 were selected does not imply that the other channels have no predictive power {or are uninformative. Rather, in the context
of penalized regression, coefficients are selected and estimated conditional on the variables already selected
in the model. As such, the absence of significance for other channels indicates that they do not
provide additional predictive power beyond what is already captured by channel 7.
This consideration directly motivates our use of the elastic net framework: Beyond predictive
performance, our objective is to identify a parsimonious set of predictors. From an operational
perspective, it is desirable to rely on as few channels as possible — ideally a single one — while
maintaining good predictive performance. In this sense, the prominence of channel 7 reflects its
role as the strongest standalone predictor within the available set.} 

{Although the elastic net model identifies channel 7 as the strongest standalone channel, examining channel-specific models remains informative for understanding the individual contribution and potential utility of each channel under different practical constraints, for instance, in situation where channel 7 data are unavailable, unreliable or where robustness across other channels is required.}
One may therefore still  wonder how the results of the elastic net regularized  model with all 8 channels (and 10 lags for each) compare against channel-specific models that only include the lags of a particular channel under investigation.
A detailed overview of the confusion matrices and classification reports for all eight channel-specific models are provided in Appendix \ref{app:channels}.

This channel-specific analysis confirms Channel 7 to be the strongest overall performer, it has an $F_1$-score for class 0 of 0.9607 and an $F_1$-score for class 1 of 0.5815; close to the results of Table \ref{tab:classification-report}. The overall accuracy is 0.9281. It retains all eleven features in the model, demonstrating both robustness and stability. Its high accuracy and minimal compromise in detecting non-flare events make it the most reliable single-channel predictor.

Channels 4 and 5 also show strong performance. Both achieve high $F_1$-scores for class 0 (non-flare events), with scores of 0.9703 and 0.9678, respectively.  Channel 4 excels in precision (0.8955) while still maintaining moderate recall (0.2801), resulting in a meaningful $F_1$-score of 0.4267 for flares. Channel 5 offers slightly lower precision (0.6859) but considerably improves recall to 0.3536, producing a higher $F_1$-score of 0.4667. 

In contrast, Channel 6 prioritizes recall above all, achieving a solar flare recall of 0.7363, the highest among all channels, but at the cost of precision (0.2497) and overall accuracy (81.39\%). However, the $F_1$-score of class 0 is still very high with a value of 0.8907. This makes it particularly useful in high-risk environments where missing a solar flare is more critical than issuing false alarms.

Channels 1, 2, and 3 perform considerably worse, with recall values for the flare class well below 10\% in Channels 2 and 3, and only a modest improvement in Channel 1. Their low $F_1$-scores and sparse detection of flares suggest limited standalone utility.

Finally, Channel 8 provides competitive recall (0.7481) similar to Channel 6, but with even lower precision (0.2153). While it contributes significantly to flare detection, the large number of false positives reduces its reliability in isolation.

In summary, while all channels demonstrate some predictive ability, Channel 7 stands out as the most effective overall, offering strong performance across both classes without extreme sacrifices in either direction. Channels 4, 5, 6, and 8 also show notable strengths depending on the desired balance between precision and recall.

\section{Discussion}
\label{sec:dis}
\subsection{Physical interpretation}

In the results above, we show that especially the higher frequency channels perform well as predictors of strong (M5+) solar flares in X-ray. Channels 6, 7 and 8 correspond to the frequencies of 4.995 GHz, 8.8 GHz and 15.400 GHz, respectively. 

This it to be expected. It has been shown by \citet{Giersch2017} that SRBs (including those associated with flares) are, in contrast to the results above, mostly expected at lower frequency channels, predominantly below 1 GHz. As also discussed in Section \ref{sec:int}, especially at the lower frequencies, we expect the corresponding radio emission not to be produced by relativistic electrons, but instead by plasma waves. These signatures correspond to all solar radio signatures, thus also those originating from weak solar flares and other space weather phenomena, such as coronal mass ejections, which generally occur much more frequently than the M5+ flares that we are investigating here.  As a result, it can be expected that most of the signals at channels 1, 2 and 3 (245 MHz, 410 MHz and 610 MHz) would originate from events not interesting from the perspective of this study, which would explain their poor performance as predictors. 

On the other hand, the higher frequency channels perform much better for these strong events. As explained in Section \ref{sec:int}, centimeter to decimeter wavelengths may be generated by non-thermal electron populations producing gyrosynchrotron emission, which also correlates well with their hard X-ray flux \citep{Dennis1988} generated by the same electron population, and thus, by extension also with their soft X-ray flux \citep{Shamsutdinova2024}. Thus, a strong increase in microwave emission generally corresponds to a strong increase in soft X-ray, which is what we likely see at these channels {6, 7 and 8}.

{The time shift that we observe between the radio emission and the soft X-ray emission is also to be expected and has been shown in other studies (see, for instance \citealp{Isola2007}). It can be attributed to the different processes taking place during flaring producing different emission \citep{Fletcher2011}. The flaring event generally starts with an impulsive stage that typically lasts up to a few minutes, during which accelerated electrons and ions interact with chromospheric plasma and create hard X-ray and millimeter-range radio waves \citep{Brown1971, Hudson1972}. Higher energetic particles can penetrate deeper into the solar atmosphere, which also means that the various radio signatures can be used to probe deeper solar atmospheric layers, see, for instance, the recent work of \cite{Ferrente2024}. Afterwards, in the gradual stage, when the plasma is cooling down, soft X-ray is produced. The peak in hard X-ray and in millimeter radio would thus indeed be expected to precede the soft X-ray peak by a few minutes, as shown in our results.}

As reported in Section \ref{sec:res}, there have also been a few strong solar flares without an associated radio signature in the eight observed channels. \citet{Benz2007} suggests that all solar flares stronger than C5 should produce a radio signal when observed radially above the source. As already discussed, the fact that no signature was observed may have been due to the fact that the radio signature took place at frequencies between the observed channels or completely outside of the range of frequencies investigated in this work. Alternatively, as also discussed by \citet{Benz2007}, these flares may have been close to the limb of the Sun.

\subsection{Limitations and future work}
As shown in Section \ref{sec:res}, the current best model has a nearly perfect detection of true negatives (no solar flare) but detects only about two thirds of actual solar flares. 
Some of these scores could  potentially be improved by including larger time windows/more lagged series, but that may also considerably increase the complexity of the model or sensitivity to noise. Our model also currently uses a fixed value of $\alpha =0.5$ for the elastic net penalty, representing an equal balance between $\ell_1$ (lasso) and $\ell_2$ (ridge) penalization. While this choice of $\alpha$ generally offers a balanced trade-off between sparsity and stability, it may not be optimal for all datasets, and other $\alpha$s may have to be explored in the future to further improve the performance of the model. The current model is also based on a purely linear logistic regression and overlooks possible nonlinear effects.

Another limitation concerns the availability and quality of radio data. Many observations had to be discarded due to highly noisy or otherwise corrupted data (e.g., giving negative amplitudes or showing signs of {radio} interference), or because {radio} data was simply not available at the time of the day. Combining the current data with sources external to RSTN, {such as those in e-Callisto \citep{Benz2005Callisto}}, would, however, be very difficult as these typically use other frequencies and have completely different specifications such as sensitivity and similar.

Operationally speaking, it should also be  noted that for some applications, the comparison of the estimated probability $\hat{\pi}_{t}$ against the threshold $0.5$ as an operational threshold for decision may be insufficient. In critical applications, a lower threshold may be favorable to prioritize true positives and minimize the risk of missing critical events. Here, we also only consider events of class M5+, as these are generally considered to be severe enough to substantially influence HF operations in the Netherlands. However, solar flare-related HF absorption loss is latitude-dependent (strongest in the sub-solar point), meaning that at lower latitudes, even weaker flares may create significant disruptions. This motivates a multiclass framework in which multiple solar flare categories are modeled instead of using a single binary threshold in a future study.


\section{Conclusion}
\label{sec:con}
In this work, we investigated the potential of using radio channels of the Radio Solar Telescope Network (RSTN) to predict whether a strong (M5+ class) X-ray solar flare is taking place. The motivation behind this research is the fact that the current technology used for monitoring and classification of solar flares, the X-ray sensor aboard the Geostationary Operational Environmental Satellites (GOES), is based in space and thus also vulnerable to strong space weather effects. In case this space asset becomes damaged or otherwise unavailable, we need to have other indicators of strong solar activity. 

We used RSTN radio data and GOES-16 X-ray data from a span of two years to develop an elastic net regularized logistic regression model, balancing model complexity and its $F_1$ score (which represents a harmonic mean of precision and recall). Having noticed that the radio signatures oftentimes appeared a few minutes before the M5 threshold of the X-ray flux was reached, ``lagged" time series of up to 10 minutes were used in the model. 

The model achieved an overall accuracy of 93.05\% and demonstrated that especially at the higher frequency channels, radio signatures can be most frequently expected up to {3 to 4} 
minutes before the X-ray flux exceeds the M5 threshold, meaning that these observations may even be useful for issuing late warnings. While the model performed very well on the non-flare events, with a high precision (0.9716) and recall (0.9527), the precision and recall for flare predictions were 0.5303 and 0.6568, meaning that one third of strong solar flares was not detected. This shows the difficulty of achieving a balance to accurately predict rare occurrences. 

The results of the eight RSTN channels showed that especially the higher frequencies (channel 6 at 4.995 GHz, channel 7 at 8.8 GHz and channel 8 at 15.4 GHz) carry predictive power. Channel 7, at 8.8 GHz, was the most reliable. While we would generally expect more radio bursts to occur at lower frequencies \citep{Giersch2017}, the radio signatures at these frequencies are typically caused by all kinds of space weather events, predominantly by weaker solar flares, coronal mass ejections et cetera. This explains why these lower frequency channels (245 MHz, 410 MHz and 610 MHz) generated too many false positives and were thus not found to be a reliable predictor of strong solar flares. As has been suggested \citep{Benz2007}, strong X-ray may lead also to strong microwave radiation, explaining why for stronger flares, we would also expect signatures at these higher channels. The channels 4 and 5 in between at 1.415 GHz and 2.695 GHz offer strong precision and moderate recall. It is thus emphasized that while channel 7 offers the best standalone performance, other channels may be prioritized depending on the specific application.

The analysis of false classifications showed that the most common reason for misidentification was the timing of the radio signatures, which sometimes appeared well outside of the considered window. Increasing the window sizes could help improve the model accuracy, but it would also increase the model complexity and possibly its sensitivity to noise. Additional possibilities to expand the model would be to further experiment with the balance between the lasso and ridge penalties in the model and considering models that allow for nonlinearities.

Despite these limitations, the present study shows that monitoring radio signatures especially at the microwave wavelengths may be a powerful alternative to monitor M5+ class solar flares in case there is no access to space-based X-ray measurements. {In addition, the best performing channels indicate that the respective radio signatures more often than not appear approximately a few minutes before the X-ray threshold is exceeded. This is consistent with what would be expected based on our current understanding of the flaring process, and it can be especially useful for automated solar radio monitoring systems to issue warnings to the users of long-distance radio systems.} A multiclass framework may have to be developed  for the model to be applicable globally, since especially at lower latitudes, even weaker solar flares may produce significant enough effects. 

\section{Acknowledgments}
We thank the editor, and referees for their constructive and detailed comments which substantially improved the quality of the manuscript. 
The authors would like to thank C. Marqué from Solar-Terrestrial Center of Excellence, Royal Observatory of Belgium for useful discussions. 

\section{Funding}

This research was financially supported by the Dutch Research Council (NWO) under the grant number VI.Vidi.211.032. 

\section{Data availability}

{The RSTN radio datasets used in this study were obtained from the online public NCEI repository AFRL Products and Data available on \url{https://www.ncei.noaa.gov/products/space-weather/partners/arfl-products-data}. The X-ray datasets from GOES-16 were obtained from the NOAA online public repository GOES-R Extreme Ultraviolet and X-ray Irradiance Sensors (EXIS) available on \url{https://www.ncei.noaa.gov/products/goes-r-extreme-ultraviolet-xray-irradiance}.}

\bibliography{jswsc}

@article{Benz2005,
  author = {Benz, Arnold O. and Grigis, Paolo C. and Csillaghy, André and Saint-Hilaire, Pascal},
  title = {Survey on Solar X-ray Flares and Associated Coherent Radio Emissions},
  journal = {Solar Physics},
  volume = {226},
  number = {1},
  pages = {121--142},
  year = {2005},
  doi = {10.1007/s11207-005-5254-5}
}

@ARTICLE{Isola2007,
       author = {{Isola}, C. and {Favata}, F. and {Micela}, G. and {Hudson}, H.~S.},
        title = "{The correlation between soft and hard X-rays component in flares: from the Sun to the stars}",
      journal = {Astronomy \& Astrophysics},
         year = 2007,
        month = sep,
       volume = {472},
       number = {1},
        pages = {261-268},
          doi = {10.1051/0004-6361:20077643},
archivePrefix = {arXiv},
       eprint = {0707.2322},
 primaryClass = {astro-ph},
       adsurl = {https://ui.adsabs.harvard.edu/abs/2007A&A...472..261I}
}

@ARTICLE{Fletcher2011,
       author = {{Fletcher}, L. and {Dennis}, B.~R. and {Hudson}, H.~S. and {Krucker}, S. and {Phillips}, K. and {Veronig}, A. and {Battaglia}, M. and {Bone}, L. and {Caspi}, A. and {Chen}, Q. and {Gallagher}, P. and {Grigis}, P.~T. and {Ji}, H. and {Liu}, W. and {Milligan}, R.~O. and {Temmer}, M.},
        title = "{An Observational Overview of Solar Flares}",
      journal = {Space Science Reviews},
         year = 2011,
        month = sep,
       volume = {159},
       number = {1-4},
        pages = {19-106},
          doi = {10.1007/s11214-010-9701-8},
archivePrefix = {arXiv},
       eprint = {1109.5932},
 primaryClass = {astro-ph.SR},
       adsurl = {https://ui.adsabs.harvard.edu/abs/2011SSRv..159...19F}
}

@ARTICLE{Ferrente2024,
       author = {{Ferrente}, F. and {Quintero Noda}, C. and {Zuccarello}, F. and {Guglielmino}, S.~L.},
        title = "{Understanding the thermal and magnetic properties of an X-class flare in the low solar atmosphere}",
      journal = {Astronomy \& Astrophysics},
         year = 2024,
        month = jun,
       volume = {686},
          eid = {A244},
        pages = {A244},
          doi = {10.1051/0004-6361/202449512},
archivePrefix = {arXiv},
       eprint = {2404.06231},
 primaryClass = {astro-ph.SR},
       adsurl = {https://ui.adsabs.harvard.edu/abs/2024A&A...686A.244F}
}

@ARTICLE{Benz2005Callisto,
       author = {{Benz}, Arnold O. and {Monstein}, Christian and {Meyer}, Hansueli},
        title = "{Callisto   A New Concept for Solar Radio Spectrometers}",
      journal = {Solar Physics},
         year = 2005,
        month = jan,
       volume = {226},
       number = {1},
        pages = {143-151},
          doi = {10.1007/s11207-005-5688-9},
archivePrefix = {arXiv},
       eprint = {astro-ph/0410437},
 primaryClass = {astro-ph},
       adsurl = {https://ui.adsabs.harvard.edu/abs/2005SoPh..226..143B}
}

@ARTICLE{Wu2024,
       author = {{Wu}, Zhao and {Kuznetsov}, Alexey and {Anfinogentov}, Sergey and {Melnikov}, Victor and {Sych}, Robert and {Wang}, Bing and {Zheng}, Ruisheng and {Kong}, Xiangliang and {Tan}, Baolin and {Ning}, Zongjun and {Chen}, Yao},
        title = "{A Multipeak Solar Flare with a High Turnover Frequency of the Gyrosynchrotron Spectra from the Loop-top Source}",
      journal = {The Astrophysical Journal},
         year = 2024,
        month = jun,
       volume = {968},
       number = {1},
          eid = {5},
        pages = {5},
          doi = {10.3847/1538-4357/ad46ff},
archivePrefix = {arXiv},
       eprint = {2405.03116},
 primaryClass = {astro-ph.SR},
       adsurl = {https://ui.adsabs.harvard.edu/abs/2024ApJ...968....5W}
}

@ARTICLE{Xu2025,
       author = {{Xu}, XiaoFeng and {Zhao}, ZhanFeng and {Liu}, Qian and {Li}, QinZheng and {Wu}, Zhao and {Lu}, Guang and {Su}, YanRui and {Chen}, Yao and {Yan}, FaBao},
        title = "{A 50-55 GHz Millimeter-wave Radiometer Spectrometer for Solar Flare Detection}",
      journal = {The Astrophysical Journal Supplement Series},
         year = 2025,
        month = jul,
       volume = {279},
       number = {1},
          eid = {29},
        pages = {29},
          doi = {10.3847/1538-4365/ade0b5},
       adsurl = {https://ui.adsabs.harvard.edu/abs/2025ApJS..279...29X}
}

@article{Giersch2017,
  author  = {Giersch, O.~D. and Kennewell, J. and Lynch, M.},
  title   = {Solar Radio Burst Statistics and Implications for Space Weather Effects},
  journal = {Space Weather},
  volume  = {15},
  pages   = {1511--1522},
  year    = {2017},
  doi     = {10.1002/2017SW001658},
  url     = {https://doi.org/10.1002/2017SW001658},
}

@ARTICLE{Dennis1988,
       author = {{Dennis}, Brian R.},
        title = "{Solar Flare Hard X-Ray Observations}",
      journal = {Solar Physics},
         year = 1988,
        month = mar,
       volume = {118},
       number = {1-2},
        pages = {49-94},
          doi = {10.1007/BF00148588},
       adsurl = {https://ui.adsabs.harvard.edu/abs/1988SoPh..118...49D}
}

@article{Shamsutdinova2024,
    author = {Shamsutdinova, J N and Kashapova, L K and Zhang, J and Reid, H and Zhdanov, D A},
    title = {Relationship between microwave and metre ranges during an impulsive solar flare},
    journal = {Monthly Notices of the Royal Astronomical Society},
    volume = {533},
    number = {2},
    pages = {1453-1462},
    year = {2024},
    month = {08},
    issn = {0035-8711},
    doi = {10.1093/mnras/stae1899},
    eprint = {https://academic.oup.com/mnras/article-pdf/533/2/1453/58883301/stae1899.pdf},
}

@ARTICLE{Reid2017,
       author = {{Reid}, Hamish A.~S. and {Vilmer}, Nicole},
        title = "{Coronal type III radio bursts and their X-ray flare and interplanetary type III counterparts}",
      journal = {Astronomy \& Astrophysics},
         year = 2017,
        month = jan,
       volume = {597},
          eid = {A77},
        pages = {A77},
          doi = {10.1051/0004-6361/201527758},
archivePrefix = {arXiv},
       eprint = {1609.04743},
 primaryClass = {astro-ph.SR},
       adsurl = {https://ui.adsabs.harvard.edu/abs/2017A&A...597A..77R}
}

@ARTICLE{Li2014,
       author = {{Li}, B. and {Cairns}, Iver H.},
        title = "{Fundamental Emission of Type III Bursts Produced in Non-Maxwellian Coronal Plasmas with Kappa-Distributed Background Particles}",
      journal = {Solar Physics},
         year = 2014,
        month = mar,
       volume = {289},
       number = {3},
        pages = {951-976},
          doi = {10.1007/s11207-013-0375-8},
       adsurl = {https://ui.adsabs.harvard.edu/abs/2014SoPh..289..951L}
}

@ARTICLE{Loughhead1957,
       author = {{Loughhead}, R.~E. and {Roberts}, J.~A. and {McCabe}, Marie K.},
        title = "{The Association of Solar Radio Bursts of Spectral Type III with Chromospheric Flares}",
      journal = {Australian Journal of Physics},
         year = 1957,
        month = dec,
       volume = {10},
        pages = {483},
          doi = {10.1071/PH570483},
       adsurl = {https://ui.adsabs.harvard.edu/abs/1957AuJPh..10..483L}
}

@ARTICLE{Bastian1998,
       author = {{Bastian}, T.~S. and {Benz}, A.~O. and {Gary}, D.~E.},
        title = "{Radio Emission from Solar Flares}",
      journal = {Annual Review of Astronomy and Astrophysics},
         year = 1998,
        month = jan,
       volume = {36},
        pages = {131-188},
          doi = {10.1146/annurev.astro.36.1.131},
       adsurl = {https://ui.adsabs.harvard.edu/abs/1998ARA&A..36..131B}
}

@ARTICLE{Garcia1998,
       author = {{Garcia}, Howard A.},
        title = "{Reconstructing the Thermal and Spatial Form of a Solar Flare from Scaling Laws and Soft X-Ray Measurements}",
      journal = {The Astrophysical Journal},
         year = 1998,
        month = sep,
       volume = {504},
       number = {2},
        pages = {1051-1066},
          doi = {10.1086/306101},
       adsurl = {https://ui.adsabs.harvard.edu/abs/1998ApJ...504.1051G}
}

@ARTICLE{Mithun2022,
       author = {{Mithun}, N.~P.~S. and {Vadawale}, Santosh V. and {Del Zanna}, Giulio and {Rao}, Yamini K. and {Joshi}, Bhuwan and {Sarkar}, Aveek and {Mondal}, Biswajit and {Janardhan}, P. and {Bhardwaj}, Anil and {Mason}, Helen E.},
        title = "{Soft X-Ray Spectral Diagnostics of Multithermal Plasma in Solar Flares with Chandrayaan-2 XSM}",
      journal = {The Astrophysical Journal},
         year = 2022,
        month = nov,
       volume = {939},
       number = {2},
          eid = {112},
        pages = {112},
          doi = {10.3847/1538-4357/ac98b4},
archivePrefix = {arXiv},
       eprint = {2210.03364},
 primaryClass = {astro-ph.SR},
       adsurl = {https://ui.adsabs.harvard.edu/abs/2022ApJ...939..112M}
}

@ARTICLE{White2011,
       author = {{White}, S.~M. and {Benz}, A.~O. and {Christe}, S. and {F{\'a}rn{\'\i}k}, F. and {Kundu}, M.~R. and {Mann}, G. and {Ning}, Z. and {Raulin}, J. -P. and {Silva-V{\'a}lio}, A.~V.~R. and {Saint-Hilaire}, P. and {Vilmer}, N. and {Warmuth}, A.},
        title = "{The Relationship Between Solar Radio and Hard X-ray Emission}",
      journal = {Space Science Reviews},
         year = 2011,
        month = sep,
       volume = {159},
       number = {1-4},
        pages = {225-261},
          doi = {10.1007/s11214-010-9708-1},
archivePrefix = {arXiv},
       eprint = {1109.6629},
 primaryClass = {astro-ph.SR},
       adsurl = {https://ui.adsabs.harvard.edu/abs/2011SSRv..159..225W}
}

@ARTICLE{Trottet1986,
       author = {{Trottet}, G.},
        title = "{Relative Timing of Hard X-Rays and Radio Emissions during the Different Phases of Solar Flares - Consequences for the Electron Acceleration}",
      journal = {Solar Physics},
         year = 1986,
        month = mar,
       volume = {104},
       number = {1},
        pages = {145-163},
          doi = {10.1007/BF00159956},
       adsurl = {https://ui.adsabs.harvard.edu/abs/1986SoPh..104..145T}
}

@ARTICLE{Ratcliffe2014,
       author = {{Ratcliffe}, H. and {Kontar}, E.~P. and {Reid}, H.~A.~S.},
        title = "{Large-scale simulations of solar type III radio bursts: flux density, drift rate, duration, and bandwidth}",
      journal = {Astronomy and Astrophysics},
         year = 2014,
        month = dec,
       volume = {572},
          eid = {A111},
        pages = {A111},
          doi = {10.1051/0004-6361/201423731},
archivePrefix = {arXiv},
       eprint = {1410.2410},
 primaryClass = {astro-ph.SR},
       adsurl = {https://ui.adsabs.harvard.edu/abs/2014A&A...572A.111R}
}

@ARTICLE{Melrose1986,
       author = {{Melrose}, D.~B. and {Cairns}, I.~H. and {Dulk}, G.~A.},
        title = "{Clumpy Langmuir waves in type III solar radio bursts}",
      journal = {Astronomy and Astrophysics},
         year = 1986,
        month = jul,
       volume = {163},
       number = {1-2},
        pages = {229-238},
       adsurl = {https://ui.adsabs.harvard.edu/abs/1986A&A...163..229M}
}

@ARTICLE{Benz2017,
       author = {{Benz}, Arnold O.},
        title = "{Flare Observations}",
      journal = {Living Reviews in Solar Physics},
         year = 2017,
        month = dec,
       volume = {14},
       number = {1},
          eid = {2},
        pages = {2},
          doi = {10.1007/s41116-016-0004-3},
       adsurl = {https://ui.adsabs.harvard.edu/abs/2017LRSP...14....2B}
}

@article{Keitarou2023,
    author = {Matsumoto, Keitarou and Masuda, Satoshi and Shimojo, Masumi and Hayakawa, Hisashi},
    title = {Relationship of peak fluxes of solar radio bursts and X-ray class of solar flares: Application to early great solar flares},
    journal = {Publications of the Astronomical Society of Japan},
    volume = {75},
    number = {6},
    pages = {1095-1104},
    year = {2023},
    month = {09},
    issn = {2053-051X},
    doi = {10.1093/pasj/psad058},
    eprint = {https://academic.oup.com/pasj/article-pdf/75/6/1095/54151707/psad058.pdf},
}

@article{Giersch2022,
author = {Giersch, O. and Kennewell, J.},
title = {Analysis of the Radio Solar Telescope Network's Noon Flux Observations Over Three Solar Cycles (1988–2020)},
journal = {Radio Science},
volume = {57},
number = {8},
pages = {e2022RS007456},
doi = {https://doi.org/10.1029/2022RS007456},
eprint = {https://agupubs.onlinelibrary.wiley.com/doi/pdf/10.1029/2022RS007456},
note = {e2022RS007456 2022RS007456},
year = {2022}
}

@ARTICLE{Hudson1972,
       author = {{Hudson}, H.~S.},
        title = "{Thick-Target Processes and White-Light Flares}",
      journal = {Solar Physics},
         year = 1972,
        month = jun,
       volume = {24},
       number = {2},
        pages = {414-428},
          doi = {10.1007/BF00153384},
       adsurl = {https://ui.adsabs.harvard.edu/abs/1972SoPh...24..414H}
}

@ARTICLE{Brown1971,
       author = {{Brown}, John C.},
        title = "{The Deduction of Energy Spectra of Non-Thermal Electrons in Flares from the Observed Dynamic Spectra of Hard X-Ray Bursts}",
      journal = {Solar Physics},
         year = 1971,
        month = jul,
       volume = {18},
       number = {3},
        pages = {489-502},
          doi = {10.1007/BF00149070},
       adsurl = {https://ui.adsabs.harvard.edu/abs/1971SoPh...18..489B}
}

@online{DISTURB,
  author = {{Defensie}},
  title = {Instrument Disturb bepaalt: zon of vijand},
  year = {2025},
note = {},
  url = {https://www.defensie.nl/actueel/nieuws/2024/05/30/instrument-disturb-bepaalt-zon-of-vijand}
}

@article{Kane1981,
  author = {Kane, S. R.},
  title = {Energetic electrons, Type III radio bursts, and impulsive solar flare X-rays},
  journal = {The Astrophysical Journal},
  volume = {247},
  pages = {1113},
  year = {1981},
  doi = {10.1086/159121}
}

@article{Woods2024,
  author = {Woods, T. N. and Eden, T. and Eparvier, F. G. and Jones, A. R. and Woodraska, D. L. and Chamberlin, P. C. and Machol, J. L.},
  title = {GOES‐R Series X‐Ray Sensor (XRS): 1. Design and Pre‐Flight Calibration},
  journal = {Journal of Geophysical Research: Space Physics},
  volume = {129},
  number = {11},
  year = {2024},
  doi = {10.1029/2024JA032925}
}

@article{ZouHastie2005,
  author = {Zou, H. and Hastie, T.},
  title = {Regularization and variable selection via the elastic net},
  journal = {Journal of the Royal Statistical Society: Series B (Statistical Methodology)},
  volume = {67},
  number = {2},
  pages = {301--320},
  year = {2005},
  doi = {10.1111/j.1467-9868.2005.00503.x}
}

@ARTICLE{Benz2007,
       author = {{Benz}, Arnold O. and {Braj{\v{s}}a}, Roman and {Magdaleni{\'c}}, Jasmina},
        title = "{Are There Radio-quiet Solar Flares?}",
      journal = {Solar Physics},
         year = 2007,
        month = feb,
       volume = {240},
       number = {2},
        pages = {263-270},
          doi = {10.1007/s11207-007-0365-9},
archivePrefix = {arXiv},
       eprint = {astro-ph/0701570},
 primaryClass = {astro-ph},
       adsurl = {https://ui.adsabs.harvard.edu/abs/2007SoPh..240..263B}
}

@book{hastie2009elements,
  title={The Elements of Statistical Learning: Data Mining, Inference, and Prediction},
  author={Hastie, Trevor and Tibshirani, Robert and Friedman, Jerome H.},
  edition={2},
  year={2009},
  publisher={Springer},
  address={New York, NY, USA},
  isbn={Hardcover: 978-0-387-84857-0, ISBN eBook: 978-0-387-84858-7},
  doi={10.1007/978-0-387-84858-7}
}

@article{belloni2016post,
  title={Post-selection inference for generalized linear models with many controls},
  author={Belloni, Alexandre and Chernozhukov, Victor and Wei, Ying},
  journal={Journal of Business \& Economic Statistics},
  volume={34},
  number={4},
  pages={606--619},
  year={2016},
  publisher={Taylor \& Francis}
}

@article{belloni2017program,
  title={Program evaluation and causal inference with high-dimensional data},
  author={Belloni, Alexandre and Chernozhukov, Victor and Fern{\'a}ndez-Val, Iv{\'a}n and Hansen, Christian},
  journal={Econometrica},
  volume={85},
  number={1},
  pages={233--298},
  year={2017},
  publisher={Wiley Online Library}
}

\newpage
\begin{appendix}


\section{Pre-analysis and statistics}

\begin{figure}[h!]
    \centering
\includegraphics[width=\linewidth]{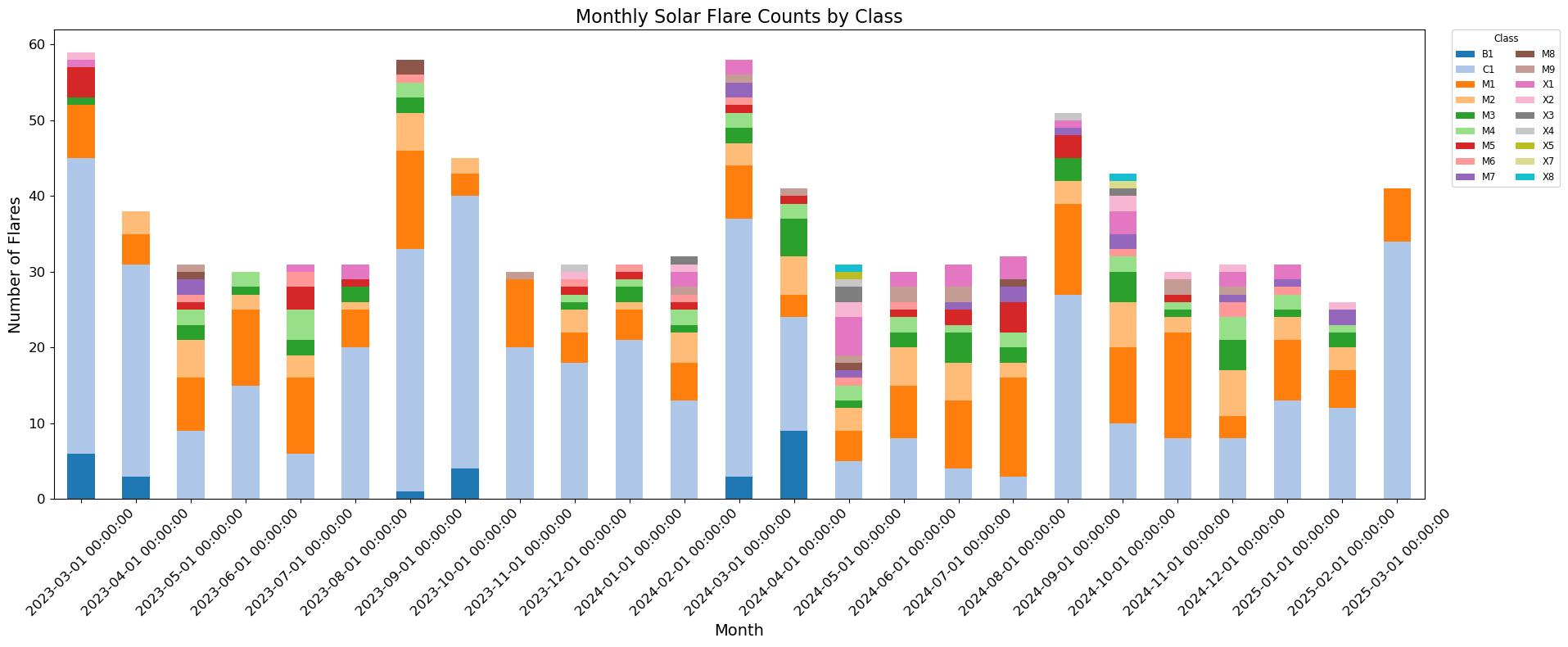}
\vspace{0.5cm}

\includegraphics[width=\linewidth]
{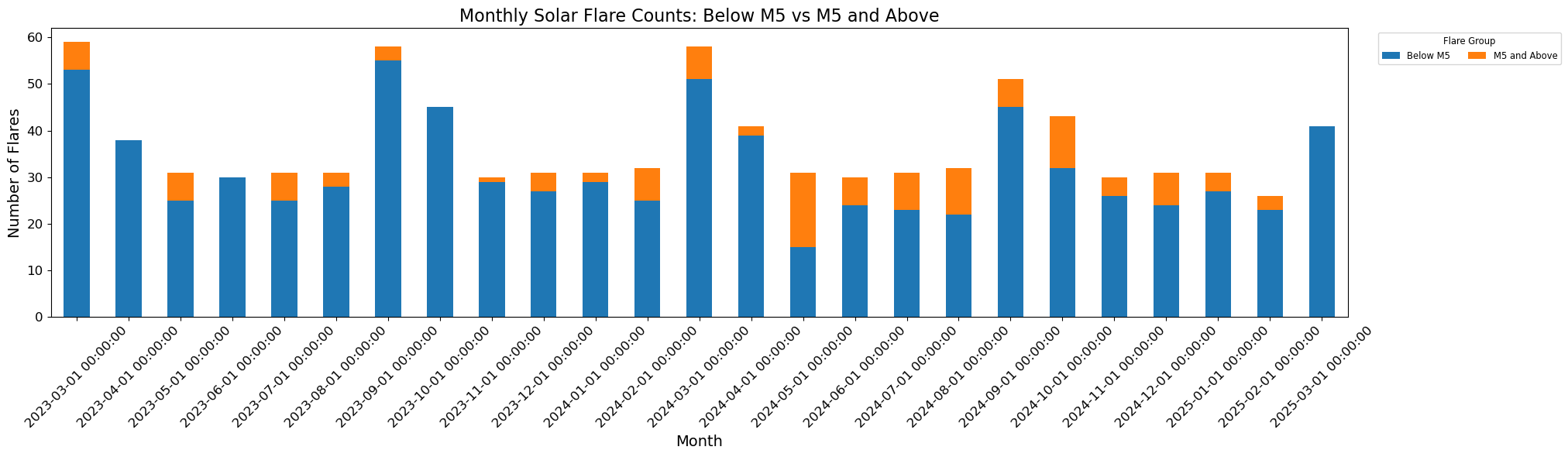}
    \caption{Monthly solar flare count by class (top) and monthly solar flare counts, below versus above  M5, (bottom) from March 2023 to March 2025.}
    \label{fig:flares_all_Xray}
\end{figure}

\begin{figure}
    \centering
    \includegraphics[width=1\linewidth]{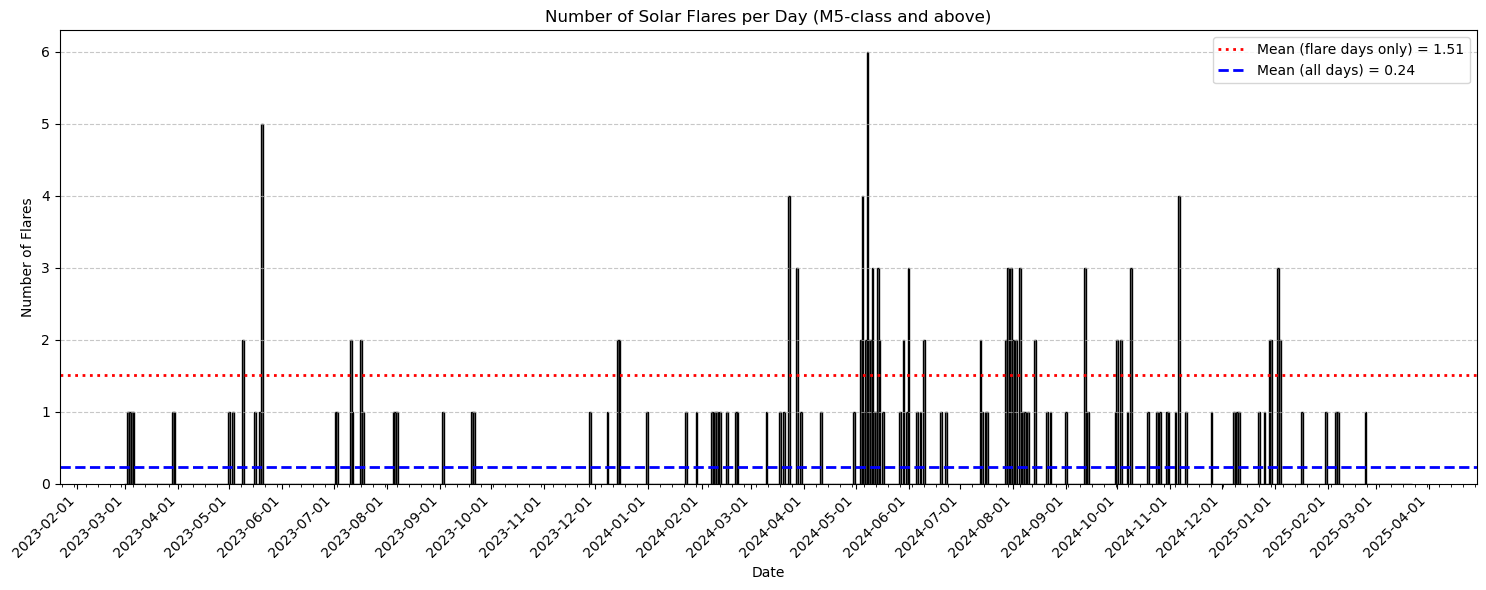}
    \caption{Number of solar flares per day (M5-class and above) from March 2023 to March 2025.}
    \label{fig:flares_per_day}
\end{figure}

\begin{figure}[h!]
    \centering
    \includegraphics[width=1\linewidth]{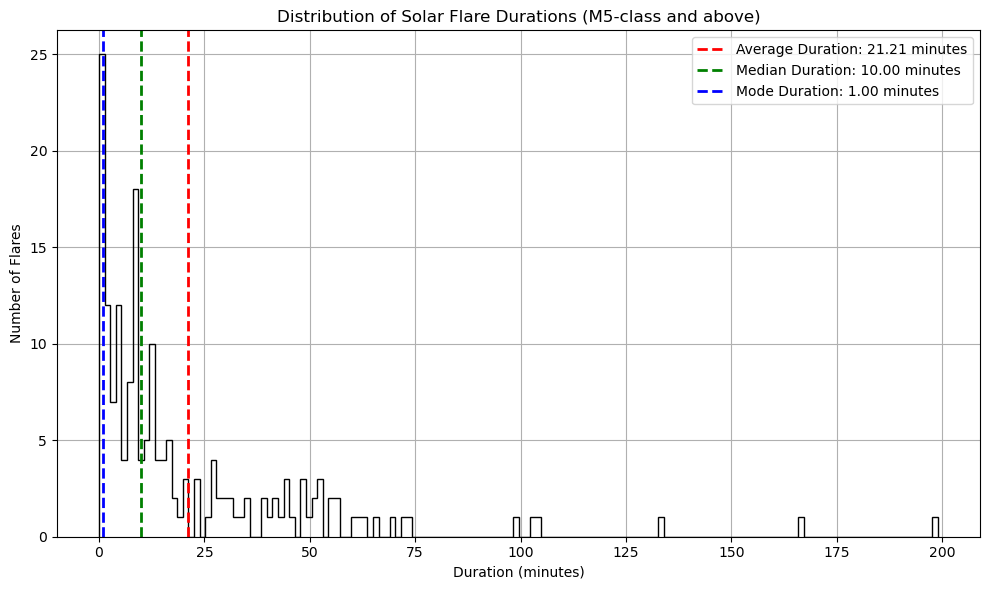}
    \caption{Distribution of solar flare durations (M5-class and above) from March 2023 to March 2025.}
    \label{fig:flare_duration}
\end{figure}

\newpage
\clearpage
\section{Pre-processing}

\begin{figure}[h!]
    \centering
    \includegraphics[width=0.9\linewidth]{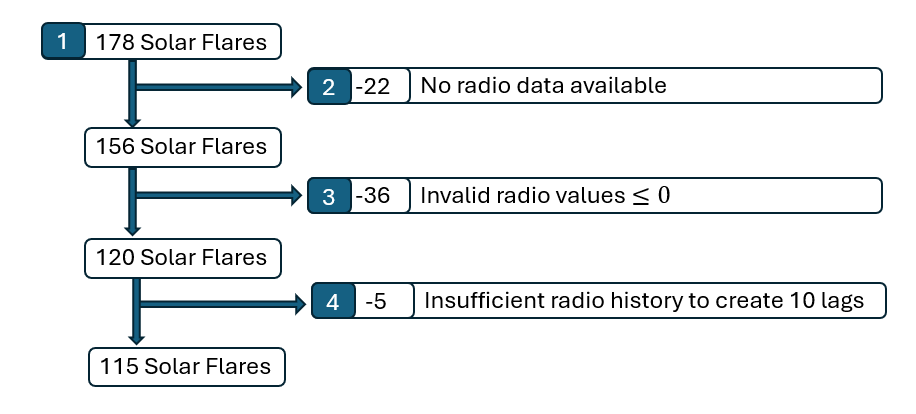}
    \caption{Pipeline of data pre-processing steps.}
    \label{fig:figure data selectio}
\end{figure}

\newpage
\clearpage

\newpage
\clearpage

\section{Time-series cross-validation to tune $\lambda$} 

We use time-series cross-validation to tune the parameter $\lambda$, that regulates the degree of sparsity in the elastic net regularized logistic regression model. 
We use time-series cross-validation, not regular cross-validation,  to preserve the temporal ordering of the data.

\subsection{The procedure to select $\lambda$} \label{app:lambda}

We start by constructing 
a grid of 10 logarithmically spaced values of $\lambda$, ranging from 10$^{-2}$ to 10$^{5}$, to  explore the spectrum of model complexity from including to excluding all radio signature variables.
We then use time-series cross-validation  to select $\lambda$. In particular, the first 41\,251 one-minute observations (approximately 70\%, the split was done at the boundary of a window with solar flare activity, see Section \ref{subsec:data}) are used for training and tuning, while the remaining (approximately 30\%) serve as a test sample to evaluate the accuracy of the models on new, out-of-sample data in Section \ref{sec:res}.
We then split the first 41\,251 observations into 5 consecutive folds of roughly equal size, see Table \ref{tab:tscv_folds}, to tune $\lambda$.
We estimate the elastic net regularized logistic models for each of the 10 values of $\lambda$ on a particular training fold, then use the consecutive tuning fold to compute the $F_1$ cross-validation score for each $\lambda$ value in the grid. We repeat this procedure five times, for each fold,  and then average the $F_1$ cross-validation scores across the five folds, for each value of $\lambda$ to finally select that value of $\lambda$ resulting in the best average cross-validation score. 
Once the optimal $\lambda$ is selected, we train the final elastic net regularized regression model on the whole training and tuning data (i.e.\ the first 41\,251 observations) before evaluating the performance of the model on the test data.

\begin{table}[h]
\centering
\caption{Set-up of time-series cross-validation procedure to tune $\lambda$.}
\label{tab:tscv_folds}
\begin{tabular}{c l l}
\hline
\textbf{Fold} & \textbf{Training observations}        & \textbf{Tuning observations}           \\
\hline
1            & 0\,–\,6\,874                  & 6\,875\,–\,13\,749           \\
2            & 0\,–\,13\,749                 & 13\,750\,–\,20\,624          \\
3            & 0\,–\,20\,624                 & 20\,625\,–\,27\,499          \\
4            & 0\,–\,27\,499                 & 27\,500\,–\,34\,374          \\
5            & 0\,–\,34\,374                 & 34\,375\,–\,41\,251          \\
\hline
\end{tabular}
\end{table}

When estimating the logistic regression models, it is important to account for the class imbalance. From the 41\,251 observations in the training and tuning set, 39\,160 (roughly 95\%) correspond to solar flare inactivity and only 2\,092 one-minute observations correspond to solar flare activity (5.07\%). Such an imbalance can bias the model toward the majority class. To correct for this and help balance the influence of both classes during model training, we introduce class weights in the loss function, thereby giving more weight to the minority class. We set the weights inversely proportional to class frequencies, namely 1/$p_0$ for time points without solar flare activity , with $p_0$ the proportion of non-solar flare observations and 1/$p_1$ for time points with  solar flare activity, with $p_1$ the proportion of solar flare observations.

\subsection{Results of the hyperparameter search} \label{app:lambda:results}

\begin{figure}
    \centering
\includegraphics[width=0.67\linewidth]{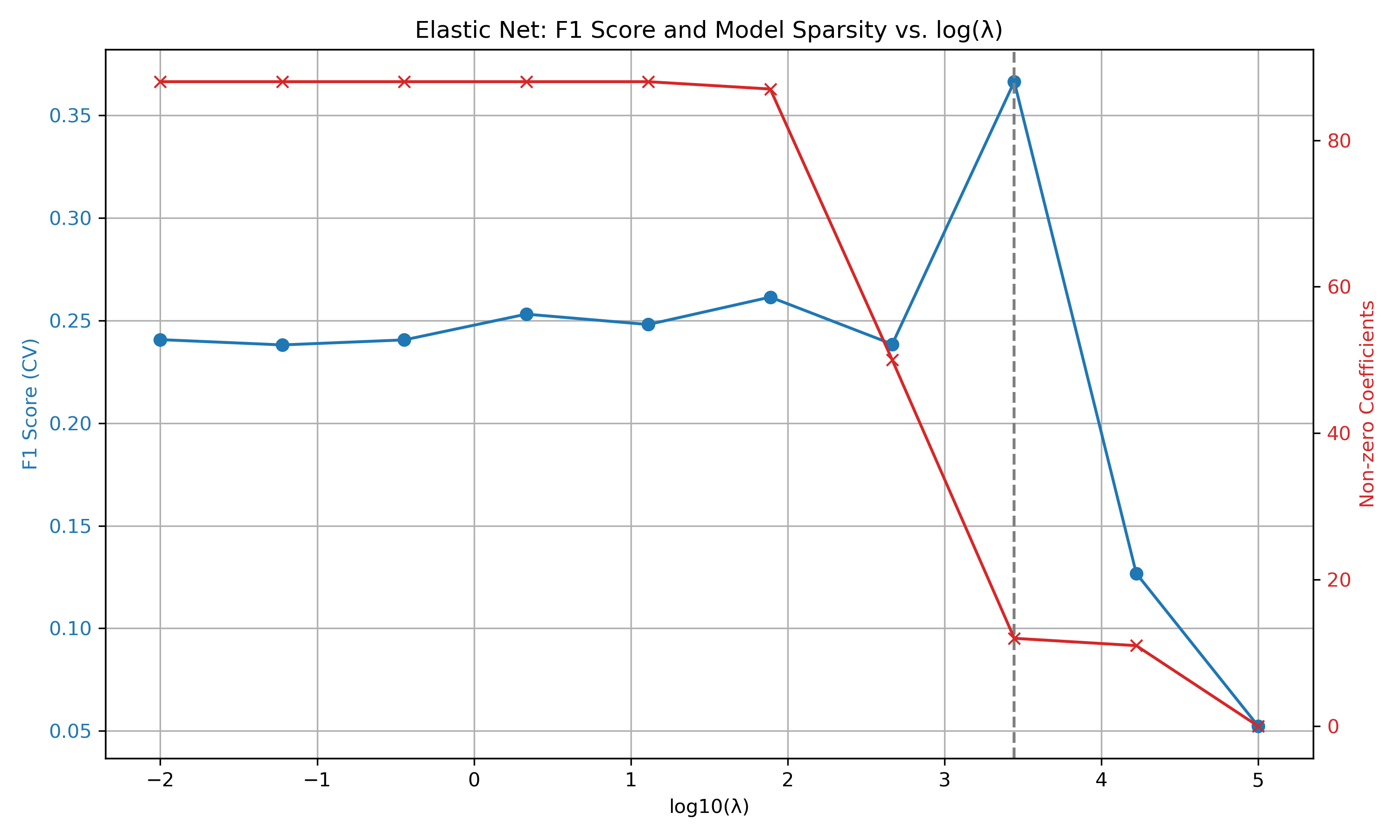}

\includegraphics[width=0.67\linewidth]{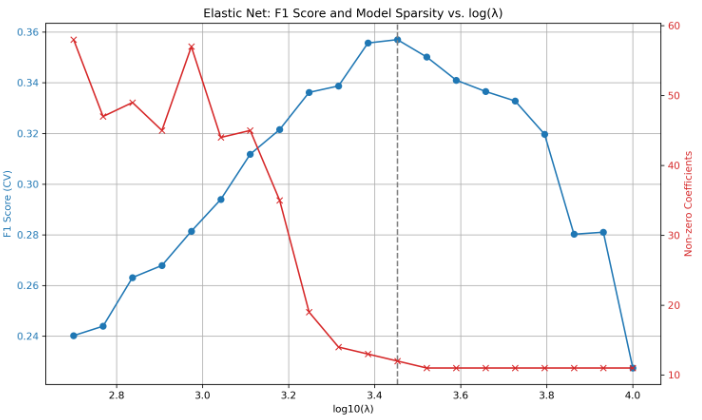}
    \caption{Initial (top) and refined (bottom) elastic net hyperparameter tuning search for $\lambda$.}
 \label{fig:hyperparametertuning}
\end{figure}

Figure \ref{fig:hyperparametertuning}, top panel, shows the results of the hyperparameter search detailed in Section \ref{app:lambda}.
The blue curve represents the mean cross-validated $F_1$ score and the red curve the number of non-zero coefficients, indicating model sparsity, in the corresponding elastic net regularized logistic regression model. For small values of $\lambda$, the model retains a large number of active coefficients (low sparsity), whereas at higher values, many coefficients are driven to zero (high sparsity).

The $F_1$ score improves for somewhat sparser models, with the stronger regularization helping eliminate irrelevant or noisy features. Around the peak of $F_1$ of 0.3432 at $\lambda \sim $ 2154, a second, refined search was carried out, see Figure \ref{fig:hyperparametertuning}, bottom panel. 
The refined search proceeds in exactly the same manner as the initial one but focuses on a narrower interval centered around the initially chosen $\lambda$ to refine the selection of $\lambda$.
Within this refined range, 20 new values of $\lambda$ are generated on a logarithmic scale and evaluated using the same time-series 5-fold cross-validation procedure.
From this refined search, for the peak $F_1$ score of 0.3570, the optimal $\lambda$ was found to be 2836, which we used to discuss the results in Section \ref{sec:res}.

\newpage
\clearpage

\section{Results for the channel-specific logistic regression models} \label{app:channels}
{We first report the confusion matrices and classification reports at the one-minute level across all different channels.
At the end, we summarize classification performance across all channels at the flare level.}



\begin{table}[!htbp]
  \centering
  \caption{Confusion matrix for Channel 1.}
  \label{tab:confusion-matrix-ch1}
  \begin{tabular}{lcc}
    \hline
                 & \textbf{No Predicted {Flare}} & \textbf{Predicted Flare} \\
    \hline
    \textbf{Actual: No Flare} & 16\,526               & 38                    \\
    \textbf{Actual: Flare} &  1\,231               & 115                   \\
    \hline
  \end{tabular}
\end{table}

\begin{table}[!htbp]
  \centering
  \caption{Classification report for Channel 1.}
  \label{tab:classification-report-ch1}
  \begin{tabular}{lrrrr}
    \hline
                   & \textbf{Precision} & \textbf{Recall} & \textbf{F1-score} & \textbf{Support} \\
    \hline
    0  (No Flare)            & 0.9307            & 0.9977          & 0.9630            & 16\,564            \\
    1   (Flare)           & 0.7516            & 0.0854          & 0.1534            &  1\,346            \\
    \hline
    \textbf{Accuracy} &                  &                 & \textbf{0.9291}   & \textbf{17\,910}   \\
    \hline
  \end{tabular}
\end{table}




\begin{table}[!htbp]
  \centering
  \caption{Confusion matrix for Channel 2.}
  \label{tab:confusion-matrix-ch2}
  \begin{tabular}{lcc}
    \hline
                 & \textbf{No Predicted {Flare}} & \textbf{Predicted Flare} \\
    \hline
    \textbf{Actual: No Flare} & 16\,429               & 135                   \\
    \textbf{Actual: Flare} &  1\,254               &  92                   \\
    \hline
  \end{tabular}
\end{table}

\begin{table}[!htbp]
  \centering
  \caption{Classification report for Channel 2.}
  \label{tab:classification-report-ch2}
  \begin{tabular}{lrrrr}
    \hline
                   & \textbf{Precision} & \textbf{Recall} & \textbf{F1-score} & \textbf{Support} \\
    \hline
    0   (No Flare)           & 0.9291            & 0.9918          & 0.9594            & 16\,564            \\
    1  (Flare)            & 0.4053            & 0.0684          & 0.1170            &  1\,346            \\
    \hline
    \textbf{Accuracy} &                  &                 & \textbf{0.9224}   & \textbf{17\,910}   \\
    \hline
  \end{tabular}
\end{table}



\begin{table}[h]
  \centering
  \caption{Confusion matrix for Channel 3.}
  \label{tab:confusion-matrix-ch3}
  \begin{tabular}{lcc}
    \hline
                 & \textbf{No Predicted {Flare}} & \textbf{Predicted Flare} \\
    \hline
    \textbf{Actual: No Flare} & 16\,532               & 32                    \\
    \textbf{Actual: Flare} &  1\,258               & 88                    \\
    \hline
  \end{tabular}
\end{table}

\begin{table}[!h]
  \centering
  \caption{Classification report for Channel 3.}
  \label{tab:classification-report-ch3}
  \begin{tabular}{lrrrr}
    \hline
                   & \textbf{Precision} & \textbf{Recall} & \textbf{F1-score} & \textbf{Support} \\
    \hline
    0    (No Flare)          & 0.9293            & 0.9981          & 0.9624            & 16\,564            \\
    1    (Flare)          & 0.7333            & 0.0654          & 0.1201            &  1\,346            \\
    \hline
    \textbf{Accuracy} &                  &                 & \textbf{0.9280}   & \textbf{17\,910}   \\
    \hline
  \end{tabular}
\end{table}


\begin{table}[!h]
  \centering
  \caption{Confusion matrix for Channel 4}
  \label{tab:confusion-matrix-ch4}
  \begin{tabular}{lcc}
    \hline
                 & \textbf{No Predicted {Flare}} & \textbf{Predicted Flare} \\
    \hline
    \textbf{Actual: No Flare}& 16\,520               & 44                    \\
    \textbf{Actual: Flare} &  969                & 377                   \\
    \hline
  \end{tabular}
\end{table}

\begin{table}[!h]
  \centering
  \caption{Classification report for Channel 4}
  \label{tab:classification-report-ch4}
  \begin{tabular}{lrrrr}
    \hline
                   & \textbf{Precision} & \textbf{Recall} & \textbf{F1‐score} & \textbf{Support} \\
    \hline
    0 (No Flare)             & 0.9446            & 0.9973          & 0.9703            & 16\,564            \\
    1  (Flare)            & 0.8955            & 0.2801          & 0.4267            &  1\,346            \\
    \hline
    \textbf{Accuracy} &                  &                 & \textbf{0.9434}   & \textbf{17\,910}   \\
    \hline
  \end{tabular}
\end{table}



\begin{table}[!h]
  \centering
  \caption{Confusion matrix for Channel 5.}
  \label{tab:confusion-matrix-ch5}
  \begin{tabular}{lcc}
    \hline
                 & \textbf{No Predicted {Flare}} & \textbf{Predicted Flare} \\
    \hline
    \textbf{Actual: No Flare} & 16\,346               & 218                   \\
    \textbf{Actual: Flare} &  870                & 476                   \\
    \hline
  \end{tabular}
\end{table}

\begin{table}[!h]
  \centering
  \caption{Classification report for Channel 5.}
  \label{tab:classification-report-ch5}
  \begin{tabular}{lrrrr}
    \hline
                   & \textbf{Precision} & \textbf{Recall} & \textbf{F1-score} & \textbf{Support} \\
    \hline
    0   (No Flare)           & 0.9495            & 0.9868          & 0.9678            & 16\,564            \\
    1  (Flare)            & 0.6859            & 0.3536          & 0.4667            &  1\,346            \\
    \hline
    \textbf{Accuracy} &                  &                 & \textbf{0.9393}   & \textbf{17\,910}   \\
    \hline
  \end{tabular}
\end{table}



\begin{table}[!htbp]
  \centering
  \caption{Confusion matrix for Channel 6.}
  \label{tab:confusion-matrix-ch6}
  \begin{tabular}{lcc}
    \hline
                 & \textbf{No Predicted {Flare}} & \textbf{Predicted Flare} \\
    \hline
    \textbf{Actual: No Flare} & 13\,586               & 2\,978                  \\
    \textbf{Actual: Flare} &  355                &  991                  \\
    \hline
  \end{tabular}
\end{table}

\begin{table}[!h]
  \centering
  \caption{Classification report for Channel 6.}
  \label{tab:classification-report-ch6}
  \begin{tabular}{lrrrr}
    \hline
                   & \textbf{Precision} & \textbf{Recall} & \textbf{F1-score} & \textbf{Support} \\
    \hline
    0   (No Flare)            & 0.9745            & 0.8202          & 0.8907            & 16\,564            \\
    1    (Flare)            & 0.2497            & 0.7363          & 0.3729            &  1\,346            \\
    \hline
    \textbf{Accuracy} &                  &                 & \textbf{0.8139}   & \textbf{17\,910}   \\
    \hline
  \end{tabular}
\end{table}



\begin{table}[!h]
  \centering
  \caption{Confusion matrix for Channel 7.}
  \label{tab:confusion-matrix-ch7}
  \begin{tabular}{lcc}
    \hline
                 & \textbf{No Predicted {Flare}} & \textbf{Predicted Flare} \\
    \hline
    \textbf{Actual: No Flare} & 15\,727               & 837                   \\
    \textbf{Actual: Flare} & 451                 & 895                   \\
    \hline
  \end{tabular}
\end{table}

\begin{table}[!h]
  \centering
  \caption{Classification report for Channel 7.}
  \label{tab:classification-report-ch7}
  \begin{tabular}{lrrrr}
    \hline
                   & \textbf{Precision} & \textbf{Recall} & \textbf{F1-score} & \textbf{Support} \\
    \hline
    0    (No Flare)           & 0.9721            & 0.9495          & 0.9607            & 16\,564            \\
    1    (Flare)            & 0.5167            & 0.6649          & 0.5815            &  1\,346            \\
    \hline
    \textbf{Accuracy} &                  &                 & \textbf{0.9281}   & \textbf{17\,910}   \\
    \hline
  \end{tabular}
\end{table}



\begin{table}[!h]
  \centering
  \caption{Confusion matrix for Channel 8.}
  \label{tab:confusion-matrix-ch8}
  \begin{tabular}{lcc}
    \hline
                 & \textbf{No Predicted {Flare}} & \textbf{Predicted Flare} \\
    \hline
    \textbf{Actual: No Flare} & 12\,894 & 3\,670 \\
    \textbf{Actual: Flare} &  339  & 1\,007 \\
    \hline
  \end{tabular}
\end{table}

\begin{table}[t]
  \centering
  \caption{Classification report for Channel 8.}
  \label{tab:classification-report-ch8}
  \begin{tabular}{lrrrr}
    \hline
                  & \textbf{Precision} & \textbf{Recall} & \textbf{F1-score} & \textbf{Support} \\
    \hline
    0   (No Flare)           & 0.9744            & 0.7784          & 0.8655            & 16\,564            \\
    1   (Flare)          & 0.2153            & 0.7481          & 0.3344            &  1\,346            \\
    \hline
    \textbf{Accuracy} &               &                 & \textbf{0.7762}   & \textbf{17\,910}   \\
    \hline
  \end{tabular}
\end{table}




\begin{table}[h]
\color{black}
\caption{Classification performance at the flare level: M5+ flares forecasted out of a total of 30 flares in the test set (True Positives; TP) thereby indicating how many of these were predicted to occur before, at or after the flare onset and the median positive, zero or negative lead time (in minutes) for respectively those predicted to occur before, at or after the onset. Furthermore, we report M5+ flares not forecasted (False Negatives; FN),  near misses (NM) and  $<$M5 flares falsely forecasted (False Alarms aka False Positives; FP), precision, recall and $F_1$-score. }
  \centering
\begin{tabular}{lccccccccc} 
\hline 
Channel & TP    & TPs Before/After/At & Median & FN  & NM & FA & Precision & Recall & F1-score \\
 &     &  Onset & Lead Time &  & &  &  \\ \hline 
1 & 5   & 1/1/3 & 3/0/-2 &  25    & 0  & 10  & 0.3333 & 0.1667  &  0.2222 \\ 
2 & 6   & 1/3/2   & 1/0/-1 &  24    & 0  & 10  & 0.3750  &  0.2000  & 0.2609 \\ 
3 & 3   & 0/2/1   & -/0/154 &  27    & 2  & 2   & 0.6000 &  0.1000  &  0.1714 \\ 
4 & 12  &  6/3/3  & 4.5/0/9 &  18    & 1  & 1   & 0.9231  &  0.4000  & 0.5581  \\ 
5 & 12  &  9/1/2  & 1/0/9.5&  18    & 4  & 13  & 0.4800  &  0.4000  & 0.4364 \\ 
6 & 21  &  14/3/4  & 7/0/2&  9     & 3  & 9   & 0.7000 &  0.7000 &  0.7000 \\ 
7 & 17  &  8/5/4  & 2.5/0/2 &  13    & 1  & 7    & 0.7083 &  0.5667 &  0.6296 \\  
8 & 23  & 14/3/6  & 21.5/0/2.5 &  7     & 5  & 97    & 0.1917 &  007667 & 0.3067 \\ 
\hline 
\end{tabular}
\end{table}

\end{appendix}

\end{document}